\documentclass[fleqn,usenatbib,useAMS]{mnras}

\usepackage{graphicx}	% Including figure files
\usepackage{amsmath}	% Advanced maths commands
\usepackage{amssymb}	% Extra maths symbols
\usepackage{multicol}   % Multi-column entries in tables
\usepackage{bm}		% Bold maths symbols, including upright Greek
\usepackage{pdflscape}	% Landscape pages
\usepackage{ulem}
\usepackage{tcolorbox}

\newcommand{\beq}{\begin{equation}}
\newcommand{\eeq}{\end{equation}}
\newcommand{\bse}{\begin{subequations}}
\newcommand{\ese}{\end{subequations}}
\newcommand{\bary}{\begin{eqnarray}}
\newcommand{\eary}{\end{eqnarray}}
\newcommand{\bwt}{\begin{widetext}}
\newcommand{\ewt}{\end{widetext}}

\usepackage[T1]{fontenc}
\usepackage{ae,aecompl}

\title{Very high energy emission mechanism in the extreme blazar PGC 2402248}

\author[B. Medina-Carrillo et al.]{
B.~Medina-Carrillo$^{1}$%
\thanks{Contact e-mail: \href{mailto:benjamin.medina@cinvestav.mx}{ benjamin.medina@cinvestav.mx}},\ %
Sarira Sahu$^{2}$%
\thanks{Contact e-mail: \href{mailto:sarira@nucleares.unam.mx}{sarira@nucleares.unam.mx}},\ %
G.~Sánchez-Colón$^{1}$%
\thanks{Contact e-mail: \href{mailto: gabriel.sanchez@cinvestav.mx}{gabriel.sanchez@cinvestav.mx}},\ %
Subhash Rajpoot$^{3}$%
\thanks{Contact e-mail: \href{mailto: Subhash.Rajpoot@csulb.edu}{Subhash.Rajpoot@csulb.edu}}%
\\
$^{1}$Departamento de Física Aplicada, Centro de Investigación y de Estudios Avanzados del IPN, Unidad Mérida.\\ A.P. 73, Cordemex, Mérida, Yucatán 97310, México.\\
$^{2}$Instituto de Ciencias Nucleares, Universidad Nacional Aut\'onoma de M\'exico,\\ Circuito Exterior S/N, C.U., A.P. 70-543, CDMX 04510, México.\\
$^{3}$Department of Physics and Astronomy, California State University,
1250 Bellflower Boulevard, Long Beach, CA 90840, USA.
}

\date{}

\pubyear{2022}

\begin{document}
\label{firstpage}
\pagerange{\pageref{firstpage}--\pageref{lastpage}}
\maketitle

\begin{abstract}
Extreme high-frequency peaked BL Lacs (EHBLs) are characterized by a synchrotron peak frequency exceeding $10^{17}$ Hz and a second peak that can be in the energy range of few GeVs to several TeVs. The MAGIC telescopes detected multi-TeV gamma-rays on April 19, 2018 for the first time from the EHBL PGC 2402248 which was simultaneously observed in multiwavelength by several other instruments. The broad band spectral energy distribution of the source is conventionally modelled using the leptonic and the hadronic models. Due to the success of the photohadronic model in interpreting the enigmatic very high-energy (VHE) flaring events from many high-energy blazars, we extend this model to explain the VHE events from PGC 2402248 observed by MAGIC telescopes and compare our results with other models. We conclude that the photohadronic fits are comparable and even fare better than most other models. Furthermore, we show that the spectrum is not hard and is in a low emission state. The estimated bulk Lorentz factor for this flaring event is found to be $\lesssim 34$.
\end{abstract}


%%%%%%%%%%%%%%%%%%%%%%%%%%%%%%%%%%%%%%%%%%%%%%%%%%

%%%%%%%%%%%%%%%%% BODY OF PAPER %%%%%%%%%%%%%%%%%%

% The MNRAS class isn't designed to include a table of contents, but for this document one is useful.
% I therefore have to do some kludging to make it work without masses of blank space.
%\begingroup
%\let\clearpage\relax
%\tableofcontents
%\endgroup
\newpage

\section{Introduction}
%%%%%%%
Blazars are a class of active galactic nuclei (AGN) and are the primary sources of extragalactic $\gamma$-rays~\citep{Romero:2016hjn}. The spectral energy distribution (SED) of blazars are characterized by two non-thermal peaks~\citep{Abdo_2010}. The first peak, in the infrared to X-ray energy region, is from the synchrotron photons emitted by relativistic electrons in the magnetic field of the blazar jet. The second peak, in the X-ray to $\gamma$-ray energy regime is generated either from the Synchrotron Self-Compton (SSC) scattering of high-energy electrons off the self-produced low-energy synchrotron seed photons in the propagating jet~\citep{Maraschi:1992iz,Murase_2012,Gao:2012sq} or from the scattering of the relativistic electrons off the photons from the external sources such as the accretion disk and/or the broad-line regions~\citep{Sikora:1994zb,Blazejowski:2000ck}. In both these scenarios, the low energy photons are boosted in energy by colliding with the high-energy electrons. Alternatively, the second peak can also be attributed to different combinations of leptonic and hadronic processes taking place in the blazar jet and in the surrounding environment. In the blazar family, the BL Lac objects are generally classified according to the position of the synchroton peak frequency, and are referred to as, low-energy peaked blazars (LBLs, $\nu_{peak} < 10^{14}\, \mathrm{Hz}$), intermediate-energy peaked blazars (IBLs, $10^{14}\, \mathrm{Hz} < \nu_{peak} < 10^{15}\, \mathrm{Hz}$), high-energy peaked blazars (HBLs, $10^{15}\, \mathrm{Hz} < \nu_{peak} < 10^{17}\, \mathrm{Hz}$~\citep{1995ApJ...444..567P, Abdo_2010, Boettcher:2013wxa}) and extreme high-energy peaked blazars (EHBLs, $\nu_{peak} > 10^{17}\, \mathrm{Hz}$~\citep{Costamante:2001pu}). The maximum energy of the second peak of the EHBL can reach up to a few TeV. However, due to the low flux output from the entire SED, EHBLs are difficult to detect with the current generation of the Imaging Atmospheric Cerenkov Telescopes (IACTs). Thus, the number of such detected blazers is small. Amongst all the EHBLs observed so far, 1ES 0229+200 is prototypical, with the synchroton peak at $3.5\times 10^{19}$ Hz and the SSC peak at $1.5\times 10^{27}\, \mathrm{Hz}\, (\sim 6.2$ TeV). At least two different temporal behavior have been observed among the EHBLs. Some of them constantly exhibit extreme properties and variability timescales ranging from months to years, with the possibility of small amplitude variability~\citep{Aharonian:2007nq,Costamante:2017xqg}. Examples of these sources are 1ES 0229+200, 1ES 0347-232, RGB J0710+591, and 1ES 1101-232~\citep{Aharonian:2007nq,Costamante:2017xqg}. Others are temporary members of the EHBL family, such as Markarian 421 (Mrk 421), Markarian 501 (Mrk 501) and 1ES 1959+650~\citep{refId0,Foffano:2019itc}, and are well known HBLs but sometimes show the extreme behavior. It is believed that the EHBL class might be a complex population of sources having different spectral behaviors at VHE ($>\, 100$ GeV) $\gamma$-rays and with different subclasses within~\citep{Foffano:2019itc}. In fact, recently, we have shown that a temporary EHBL may envisage two different subclasses~\citep{Sahu:2020tko, Sahu:2021wue, Sahu:2020kce, 10.1093/mnras/stac2093}.

The VHE $\gamma$-rays observed by the Cerenkov telescopes from the extragalactic sources are attenuated by the extragalactic background light (EBL) through $e^+e^-$--pair production~\citep{1992ApJ...390L..49S,doi:10.1126/science.1227160} and through interactions with the background radiation at various electromagnetic frequencies. As such, the detection of high-energy neutrinos and photons from sources at cosmological distances (redshift $z > 0.1$) requires a detailed account of the photon background along the line-of-sight of the source. Several EBL models~\citep{Franceschini:2008tp,Dominguez:2010bv,10.1111/j.1365-2966.2012.20841.x} have been developed to study the resulting attenuation at different redshifts. 

The shift in the second peak and the hard VHE spectra of the EHBLs are difficult to account for in the framework of the one-zone leptonic SSC model, as this model predicts a soft spectrum in the Klein-Nishina regime. Proposed solutions to overcome these problems require large values of the minimum electron Lorentz factor, the bulk Lorentz factor~\citep{refId0,refId01} and also require a small magnetic field. In an alternative scenario, it is assumed that the VHE photons are produced in the intergalactic space, rather than in the jet. The escaping ultra high energy (UHE) protons from the jet travel far from the source before interacting with the cosmic microwave background (CMB) photons and/or the EBL to produce $\gamma$-rays. Thus, these $\gamma$-rays are produced closer to the Earth and hence travel less distance compared to the $\gamma$-rays produced in the jet. As a result, the $\gamma$-rays produced from the interaction of UHE protons with the CMB or EBL suffer less absorption when arriving to the Earth~\citep{2010APh....33...81E,2011ApJ...731...51E,2013ApJ...771L..32T}. Also, many alternative models such as the two-zone leptonic model, the spine-layer structured jet model, the inverse Compton (IC) scattering of the high energy electrons with the cosmic microwave photons model, and hybrids of leptonic and hadronic models are proposed to explain these spectra~\citep{B_ttcher_2008,refId0,10.1093/mnras/stz2725}. 

Previously, we explained the VHE flaring events from many HBLs and EHBLs using the photohadronic model~\citep{Sahu_2019,10.1093/mnras/staa023}. There it was shown that the spectral index $\delta$ lies in the range $2.5 \leq \delta \le 3.0$  and the photohadronic fit can not differentiate between the HBL and EHBL. In the present work, we again use the photohadronic model to fit the VHE spectrum of PGC 2402248 and compare our results with the fittings of other aforementioned models. We conclude that our model fits are comparable and fare better than most other models.

\section{Flaring of PGC 2402248}

The MAGIC collaboration undertook an observational program to search for new EHBLs. In this program it selected the object PGC 2402248 (also known as 2WHSP J073326.7+515354) from the 2WHSP catalogue~\citep{Chang:2017} on the basis of its high synchrotron peak frequency $\nu^{peak}_{syn} =10^{17.9}\, \mathrm{Hz}$. The MAGIC telescopes observed the source from January 23 to April 19, 2018 (MJD 58141-58227) for 25 nights for a total of 23.4 h. On 19th April, for the first time, it detected TeV $\gamma$-rays from the blazar PGC 2402248. During this period, simultaneous multiwavelength observations were also carried out by the KVA and the {\it Swift}-UVOT in the optical and the UV band, in the X-ray band by the {\it Swift}-XRT, and in the $\gamma$-ray band by the {\it Fermi}-LAT~\citep{10.1093/mnras/stz2725}. Observations were also carried out by the optical telescope Gran Telescopio Canarias (GTC) to estimate the redshift as it was unknown at the time. The new redshift measurement reported by the GTC is $z=0.065$~\citep{becerra2018}. The observed broad-band SED of PGC 2402248 was studied by including the $\gamma$-ray archival data collected during more than 10 years (from 4th August 2008 to 24th June 2019) by {\it Fermi}-LAT to compare the flux variability and to construct the multiwavelength SED.

The synchrotron peak frequency was estimated from the {\it Swift}-XRT data that is simultaneous to the MAGIC observations and from the non-simultaneous 105-month archival data from {\it Swift}-BAT. The newly estimated $\nu^{peak}_{syn}=10^{17.8\pm 0.3}\, \mathrm{Hz}$ was compatible with the one reported in the 2WHSP catalogue. The estimate of $\nu^{peak}_{syn}$ during different observation periods confirm that PGC 2402248 is a stationary EHBL. During the MAGIC observation period on PGC 2402248, the simultaneous observations carried out showed no significant variability except for a moderate variability in the the {\it Swift}-UVOT/XRT data. Fitting to the spectrum of the long-term observations has the possibility of averaging out the short-term variability. Although, the overall non variability in the observed SED during the MAGIC observations on PGC 2402248 shows that the source was in a stable state, the short-term flaring cannot be ignored.
    
\section{Photohadronic model}

The photohadronic model is based on the assumption of a double jet structure along the common axis~\citep{Sahu:2019lwj,Sahu_2019}. During the VHE flaring, a compact and confined smaller jet, of size $R'_f$, is formed within the bigger jet of size $R'_b$ ($R'_b > R'_f$, with the primed symbols indicating the quantity in the jet comoving frame). The photon density in the inner jet region $n'_{\gamma,f}$ is much higher than the photon density in the outer jet region $n'_{\gamma}$ ($n'_{\gamma,f} \gg n'_{\gamma}$). The photohadronic model relies on the standard interpretation of the first two peaks of the SED, i.e., the first peak is due to the synchrotron radiation of the relativistic electrons in the jet environment and the second peak is from the SSC process. Although the inner jet moves (slightly) faster than the outer jet, their respective bulk Lorentz factors satisfy $\Gamma_{in} \,> \Gamma_{ext}$, for simplicity we assume $\Gamma_{ext}\simeq \Gamma_{in}\equiv \Gamma$.

Protons are accelerated to very high energies in the inner jet region and their differential spectrum is a power-law of the form, $dN_p/dE_p \propto E^{-\alpha}_p$~\citep{Gupta_2008}, where $E_p$ is the proton energy and the spectral index $\alpha \ge 2$. In the inner jet region, the dominant process through which protons interact with the seed photons is $p+\gamma \rightarrow \Delta^+$, followed by
%%%%%%%%%%%%%%%%%%%%%%
\beq
\Delta^+\rightarrow  
\left\{ 
\begin{array}{l l}
 p\,\pi^0, & \quad \text {fraction~ 2/3}\\
  n\,\pi^+ , 
& \quad  \text {fraction~ 1/3}\\
\end{array} \right. .
\label{decaymode}
\eeq
%%%%%%%%%%%%%%%%%%%%%%
The production of $\Delta$-resonance has a cross section $\sigma_{\Delta}\sim 5\times 10^{-28}\,{\rm cm}^2$. Although the direct single pion production and the multi-pion production processes contribute, they are less efficient in the energy range under consideration here \citep{1999PASA...16..160M,2018MNRAS.481..666O}. We neglect such contributions in the present work. The produced charged and neutral pions decay through $\pi^+\rightarrow e^+{\nu}_e\nu_{\mu}{\bar\nu}_{\mu}$ and $\pi^0\rightarrow\gamma\gamma$ respectively. In the present scenario, the $\gamma$-rays produced from the neutral pion decay are the observed VHE $\gamma$-rays on Earth. 

From the $\pi^0$ decay, the observed VHE $\gamma$-ray energy $E_{\gamma}$ and the seed photon energy $\epsilon_{\gamma}$ satisfy the following condition~\citep{Sahu:2019lwj,Sahu_2019},
%%%%%%%%%%%%%%%%
\beq
E_{\gamma} \epsilon_\gamma \simeq\frac{0.032\ \Gamma{\mathcal D}}{(1+z)^{2}}\ \mathrm{GeV^2},
\label{eq:kingamma}
\eeq
%%%%%%%%%%%%%%%%
where ${\cal D}$ is the Doppler factor. The observed $\gamma$-ray energy $E_{\gamma}$ and the proton energy $E_p$ are related through 
%%%%%%%%%%%%%%%%
\beq
E_p=\frac{10\,\Gamma}{\cal D} E_{\gamma}.
\label{eq:epeg}
\eeq
%%%%%%%%%%%%%%%%
For blazars $\Gamma\simeq {\cal D}$, which gives $E_p=10\, E_{\gamma}$. For most of the VHE flaring events from the HBLs, the value of $\Gamma$ (or ${\cal D})$ is such that the seed photon energy $\epsilon_{\gamma}$ always lies in the low energy tail region of the SSC band. 

The efficiency of the $\Delta$-resonance production in the inner jet region depends on the optical depth $\tau_{p\gamma}$ and is given by
%%%%%%%%%%%%%%%%
\beq
\tau_{p\gamma}=n'_{\gamma, f} \sigma_{\Delta} R'_f,
\label{optdepth}
\eeq
%%%%%%%%%%%%%%%%
By assuming that the Eddington luminosity $L_{Edd}$ is shared equally by the jet and the counter jet during a flaring event, the luminosity $L'_{jet}$ for a seed photon of energy $\epsilon'_{\gamma}$ satisfies $L'_{jet} \ll {L_{Edd}}/{2}$ and this gives 
%%%%%%%%%%%%%%%%
\beq
\tau_{p\gamma} \ll \frac{L_{Edd}}{8\pi}\, \frac{\sigma_{\Delta}}{R'_f\,\epsilon'_{\gamma}}.
\label{optdepth2}
\eeq
%%%%%%%%%%%%%%%%
As the inner jet region is hidden, there is no direct way to determine the photon density there. Due to the adiabatic expansion of the inner jet into the outer jet, the photon density will be reduced to $n'_{\gamma}$ and the efficiency of the $p\gamma\rightarrow\Delta$ will be suppressed, leading to $\tau_{p\gamma}\ll 1$. This is the reason that in a single jet scenario the efficiency of the $\Delta$-resonance production is suppressed and one needs super-Eddington luminosity in protons to explain the observed VHE $\gamma$-ray flux in the hadronic model. However, the additional compact inner jet we consider here overcomes the excess energy budget in protons~\citep{Sahu:2016bdu}. From the SED we can calculate the photon density in the outer jet region $n'_{\gamma}$. To keep matters simple, we assume a scaling behavior of the photon densities in the inner and the outer jets as~\citep{Sahu:2019lwj,Sahu_2019}
%%%%%%%%%%%%%%%%
\beq
\frac{n'_{\gamma,f}(\epsilon_{\gamma,1})}{n'_{\gamma,f}(\epsilon_{\gamma,2})} \simeq\frac{n'_{\gamma}(\epsilon_{\gamma,1})}{n'_{\gamma}(\epsilon_{\gamma,2})}.
\label{eq:scaling}
\eeq
%%%%%%%%%%%%%%%%
\noindent
From this, we deduce that the ratio of the photon densities at two different background energies $\epsilon_{\gamma,1}$ and $\epsilon_{\gamma,2}$ in the inner (flaring) and the outer jet (non-flaring) regions are almost the same. The photon density in the outer region can be calculated in terms of the SSC photon energy $\epsilon_{\gamma}$ and its corresponding flux, $\Phi_{SSC}(\epsilon_{\gamma})$, as
%%%%%%%%%%%%%%%%
\beq
n'_{\gamma}(\epsilon_{\gamma})=\eta \left ( \frac{d_L}{R'_b} \right )^2 \frac{1}{(1+z)}\frac{\Phi_{SSC}(\epsilon_{\gamma})}{{\cal D}^{2+\kappa}\, \epsilon_{\gamma}},
\label{photondensity}
\eeq
%%%%%%%%%%%%%%%%
where the efficiency of the SSC process is defined by $\eta$ and in this work we consider 100\% efficiency by taking $\eta=1$. The luminosity distance to the source is given by $d_L$ and $\kappa=0(1)$ corresponds to a continuous (discrete) blazar jet. Using the above equation we can express $n'_{\gamma,f}$ in terms of $\Phi_{SSC}$. In many previous studies we have shown that the $\Phi_{SSC}$ in the low energy tail region is a perfect power-law for HBLs and in many EHBLs, with the synchrotron peak position always above $10^{17}$ Hz and variability timescales ranging from months to years (referred to as stationary EHBLs), it can be expressed as $\Phi_{SSC}\propto \epsilon^{\beta}_{\gamma}\propto E_{\gamma}^{-\beta}$ with $\beta \, > 0$~\citep{Sahu:2019lwj,Sahu_2019}. The observed VHE $\gamma$-ray flux $F_{\gamma}$ depends on the seed photon density $n'_{\gamma,f}$, and the high energy proton flux $F_p\equiv E^2_p\,dN/dE_p$. Expressing $n'_{\gamma,f}$ in terms of $\Phi_{SSC}$ implies that $F_{\gamma}\propto n'_{\gamma,f}\propto E^{-\beta+1}_{\gamma}$. Similarly, for $F_{\gamma}\propto F_{p}$, we deduce $F_{\gamma}\propto E^{-\alpha+2}_{\gamma}$. The VHE $\gamma$-ray flux is attenuated due to the EBL effect by a factor $e^{-\tau_{\gamma\gamma}}$, where $\tau_{\gamma\gamma}$ is the optical depth for the lepton pair production process $\gamma\gamma\rightarrow e^+e^-$ and depends on $z$ and $E_{\gamma}$. By including the EBL correction, the observed VHE flux on Earth is 
%%%%%%%%%%%%%%%%
\beq
F_{\gamma}(E_{\gamma})=F_0 \left ( \frac{E_\gamma}{TeV} \right )^{-\delta+3}\,e^{-\tau_{\gamma\gamma}}=F_{\gamma, in}(E_{\gamma})\, e^{-\tau_{\gamma\gamma}},
\label{eq:fluxgeneral}
\eeq
%%%%%%%%%%%%%%%%
where the spectral index $\delta=\alpha+\beta$ which should be in the range $2.5 \le \delta \le 3.0$~\citep{Sahu:2019lwj,Sahu_2019}. $F_0$ is the normalization constant that is fixed from the observed VHE spectrum and $F_{\gamma,in}$ is the intrinsic VHE flux. 

In the photohadronic process each pion carries 20\% of the proton energy. The decay products produced from $\pi^+$ decay, each carries approximately 25\% of the pion energy. This gives the neutrino energy $E_{\nu}=E_{\gamma}/2$ and  the neutrino flux $F_{\nu}$ can be calculated from $F_{\gamma}$, which gives~\citep{PhysRevD.87.103015}
%%%%%%%%%%%%%%%
\beq
F_{\nu}= \frac{3}{8} \,F_{\gamma}.
\label{eq:nuflux}
\eeq
%%%%%%%%%%%%%%%%%%
It is important to note that the photohadronic process works well for $E_\gamma \gtrsim 100$ GeV. However, below this energy the leptonic processes such as the electron synchrotron mechanism and the SSC process have the dominant contribution to the multiwavelength SED. As our main motivation is to interpret the VHE spectrum of the flaring event, we favor the photohadronic model for this analysis.

\section{Results and analysis}                                                                                              
Simultaneous multiwavelength observations of the blazar PGC 2042248  during the MAGIC observation period from January 23 to April 19, 2018 and the 10 year archival data collected by the Fermi-LAT telescope helped to construct the broadband SED of the object with a high degree of precision in Ref.~\citet{10.1093/mnras/stz2725}. In this reference, the X-ray data shows that the synchrotron peak frequency is in the EHBL region. Also, the observed VHE spectrum reconstructed in the energy range 0.1 TeV to 8 TeV can be described well by a power-law of the form $dN/dE_{\gamma}=f_0 (E_{\gamma}/200\  \mathrm{GeV})^{-\lambda}$ with $f_0=(1.95\pm 0.10_{stat})\times 10^{-11}\,\mathrm {ph\, cm^{-2}\, s^{-1}\, TeV^{-1}}$ and $\lambda=2.41\pm 0.17_{stat}$ and the intrinsic spectrum is also fitted with a power-law which is almost flat in this energy regime, as shown in Fig.~2 of the same reference. Also in~\citet{10.1093/mnras/stz2725}, the multiwavelength SED is fitted by using the one-zone SSC model~\citep{1992ApJ...397L...5M, Tavecchio:1998xw}, the 1D conical jet model~\citep{2015ApJ...808L..18A, 2018ApJ...861...31A}, the spine-layer model~\citep{ ref}, and the proton-synchrotron model~\citep{10.1093/mnras/stu2691}. As concluded there, extreme physical parameters would be required for the three of the four SED modelling scenarios to provide compatible models for the SED. Out of these four models the spine-one layer model fits better with theoretical predictions and provides a reasonable framework to explain the broad-band SED of PGC 2042248.

In previous studies, it was shown that the photohadronic model is successful in explaining the VHE spectra of many HBLs and EHBLs~\citep{Sahu_2019,10.1093/mnras/staa023}. It was found that the VHE spectra of the EHBLs, 1ES 0229+200, 1ES 0347-232, and several other blazars are fitted very well by taking the spectral index $\delta$ in the range $2.5\le \delta \le 3.0$. In view of the success of the photohadronic model, it is natural to extend it to the study of the VHE spectrum of the EHBL PGC 2042248.

We fit the VHE spectrum observed by the MAGIC telescopes on EHBL PGC 2042248 by using the photohadronic model and the EBL model of~\cite{Franceschini:2008tp}. The only free independent parameter in the photohadronic model is the spectral index, $\delta$, and it is estimated by varying the value of the normalization constant, $F_0$, to find the best fit values for the spectrum. The best fit is obtained for $F_0=0.7\times 10^{-12}\, \mathrm{erg\, cm^{-2}\, s^{-1}}$ and $\delta=3.0$, with $\chi^2_{min}=1.52$.

\begin{figure*}%figure 1
\includegraphics[width=5.5in]{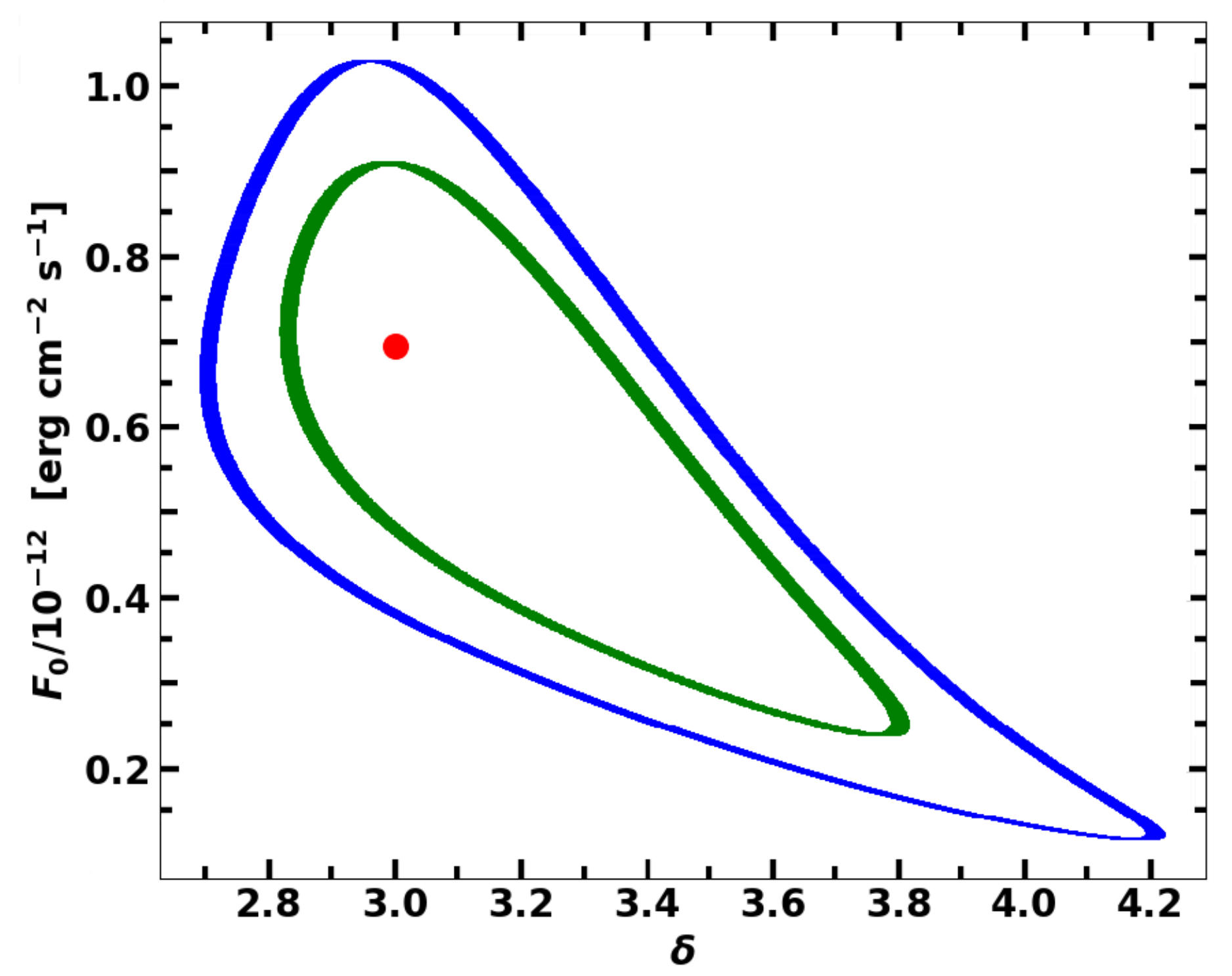}
\caption{\label{fig:figure1}Error ellipses at $\chi^2_{min}+1$ and $\chi^2_{min}+2.3$. The contour at $\chi^2_{min}+2.3$ corresponds to a coverage probability of $68.27\%$ for joint estimation of $F_0$ and $\delta$. The statistical error bars of $F_0$ and $\delta$ are obtained by varying one parameter while the other is frozen at its optimum value. The contour at $\chi^2_{min}+1$ corresponds to a coverage probability of $68.27\%$ for individual estimation of $F_0$ and $\delta$. The individual $68.27\%$ confidence intervals of $F_0$ and $\delta$ (used to build the $1\sigma$ confidence level region of Fig.~\ref{fig:figure2}) are determined by the horizontal and the vertical tangents to the ellipse, respectively.}
\end{figure*}

\begin{figure*}%figure 2
\includegraphics[width=5.5in]{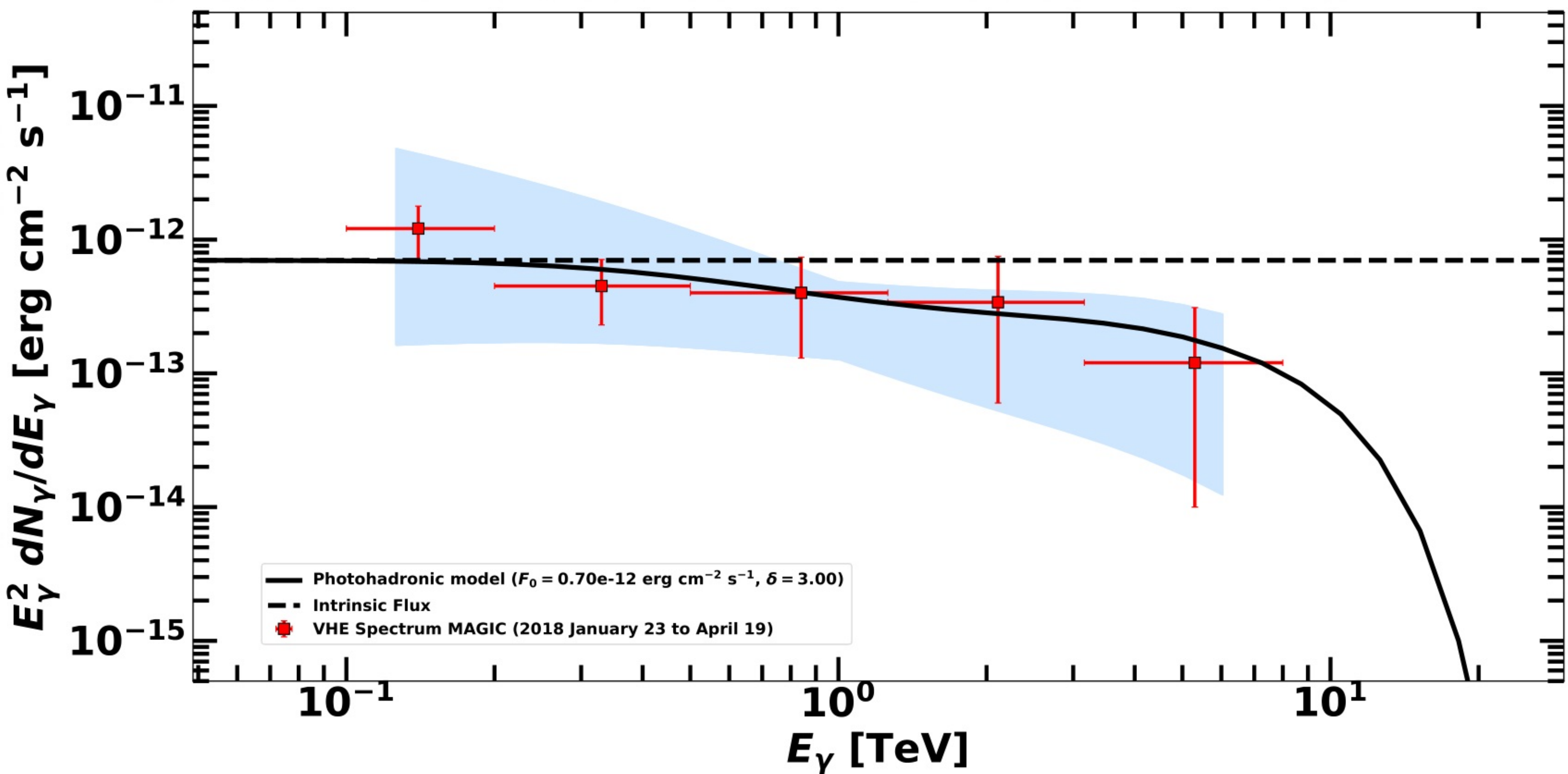}
\caption{\label{fig:figure2}The MAGIC observation of the VHE spectrum of April 19, 2018 from the EHBL PGC 2402248 is fitted with the photohadronic model using the EBL model of~\citet{Franceschini:2008tp}. The blue shaded region corresponds to the $1\sigma\ (68.27\%)$ confidence level region, obtained by varying $F_0$ and $\delta$, to their individual $68.27\%$ confidence intervals (see~Fig.~\ref{fig:figure1}). The dashed curve is the intrinsic flux.}
\end{figure*}

As defined by the error ellipse, $\chi^2_{min}+2.3$, shown in Fig.~\ref{fig:figure1}, statistical errors are obtained by varying one parameter while the other is frozen at its optimum value and they represent the $68.27\%$ confidence intervals for the joint estimation of the two parameters. The statistical errors obtained are $F_0=(0.7^{+0.33}_{-0.32})\times 10^{-12}\, \mathrm{erg\, cm^{-2}\, s^{-1}}$ and $\delta=3.0^{+0.40}_{-0.31}$. In Fig.~\ref{fig:figure2}, we show the best fit to the VHE spectrum. The shaded region corresponds to the $1\sigma\ (68.27\%)$ confidence level region and is obtained by varying $F_0$ and $\delta$ to their individual $68.27\%$ confidence intervals as defined by the ellipse $\chi^2_{min}+1$, see Fig.~\ref{fig:figure1}.

%%%%%%%%%%%%%%%%%%%%%%%%%%%%%%%%%%
%%%%%%%%%%%%%%%%%%%%%%%%%%%%%%%%%%%
%%%%%%%%%%%%%%%%%%%%%%%%%%%%%%%%%%%%%%%%%
\begin{figure*}%figure 4
\includegraphics[width=5.5in]{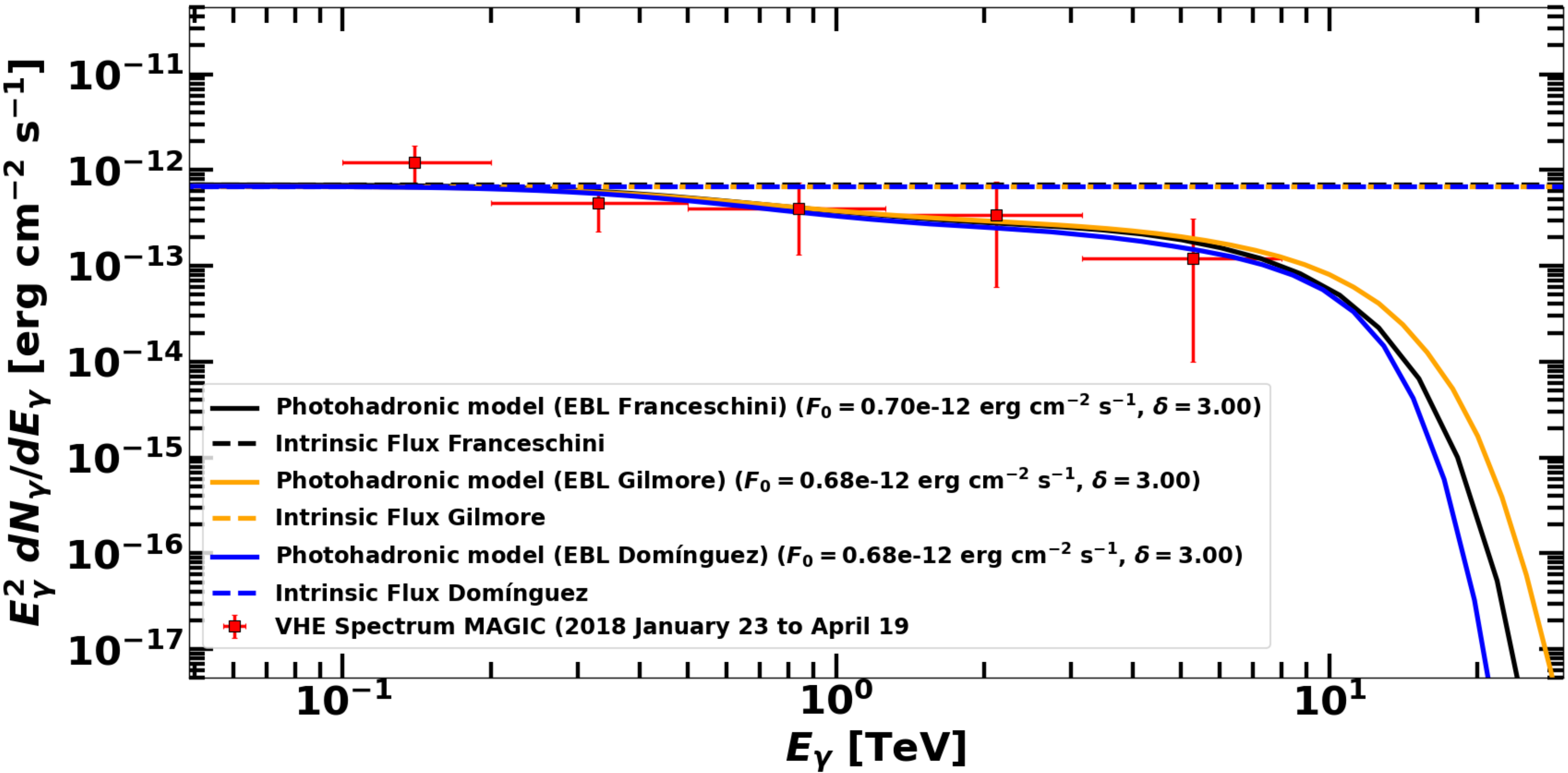}
\caption{\label{fig:figure4}The observed VHE spectrum of the EHBL PGC 2402248 is fitted with the photohadronic model by taking into account three different EBL models, by~\citet{Franceschini:2008tp},~\citet{10.1111/j.1365-2966.2012.20841.x}, and~\citet{Dominguez:2010bv}. Although there is a minor difference for $E_{\gamma} > 1$ TeV, all of them are compatible with each other. The dashed curves correspond to the intrinsic fluxes.}
\end{figure*}
%%%%%%%%%%%%%%%%%%%%%%%%%%%%%%%%%%%%%%%%%
We have also used the EBL models of~\citet{Dominguez:2010bv} and~\citet{10.1111/j.1365-2966.2012.20841.x} to fit the VHE spectrum and compared them with the photohadronic model. The results are shown in Fig.~\ref{fig:figure4}. All these models fit very well to the observed data with $\delta=3.0$ and with $F_0$ value almost the same. The comparison shows that above 1 TeV, there is a small difference in these fits. However, all these EBL models are compatible with each other. For the rest of our analysis we will use the EBL model of~\citet{Franceschini:2008tp} for comparisons purposes since the other EBL models will give similar results. 

The observed VHE spectrum of the EHBL PGC 2042248, being substantially flat, implies that its flux should increase up to several TeV with a hard spectral index~\citep{Costamante:2017xqg}. However, the photohadronic model fits very well to the VHE spectrum with $\delta=3.0$, which corresponds to a low emission state and the spectrum is soft~\citep{Sahu_2019}. The intrinsic spectrum is constant and given by $F_{\gamma}=F_0$ which is shown as dashed line in Fig.~\ref{fig:figure2} and Fig.~\ref{fig:figure4}. The spectral index of the differential proton spectrum, $\alpha=2$, is used here which corresponds to $\beta=1.0$ implying that $\Phi_{SSC}\propto E^{-1}_{\gamma}$.

%%%%%%%%%%%%%%%%%%%%%
\begin{table}
\centering
\caption{
\label{tab:tab1}In Fig.~\ref{fig:figure5} (VHE Spectrum) and in Fig.~\ref{fig:figure6} the broad-band SED of PGC 2042248 are fitted with various models: the one-zone SSC (SSC1), the 1D conical jet model (1D SSC), the spine-layer (SL), the proton synchrotron model (PS), and the photohadronic model (PH). The various parameters, like the bulk Lorentz factor ($\Gamma$), the blob Radius ($R'_b$ in units of $10^{16}$ cm~\citep{10.1093/mnras/stz2725}), and the magnetic field ($B'$ in G) used in these models are summarized below.
}
\begin{tabular}{llll}
\hline
Model & $\Gamma$ & $R^{\prime}_b$ & $B^{\prime}$\\
\hline
SSC1 & 30 & 1 & 0.01\\
1D SSC & 30 & 2.1 & 0.005\\ 
SL & 30, 5 & 3, 3.5 & 0.02, 0.1\\ 
PS & 30 & 0.1--14.6 & 1.2--46.8\\
PH &  $\le 34$ & 1 & $\sim 10^{-4.3}$\\
\hline
\end{tabular}
\end{table}
%%%%%%%%%%%%%%%%%%%%%%%%%%%

%%%%%%%%%%%%%%%%%%%%%%%%%%%%%%%%%%%%%%%
\begin{figure*}%figure 5
\includegraphics[width=5.5in]{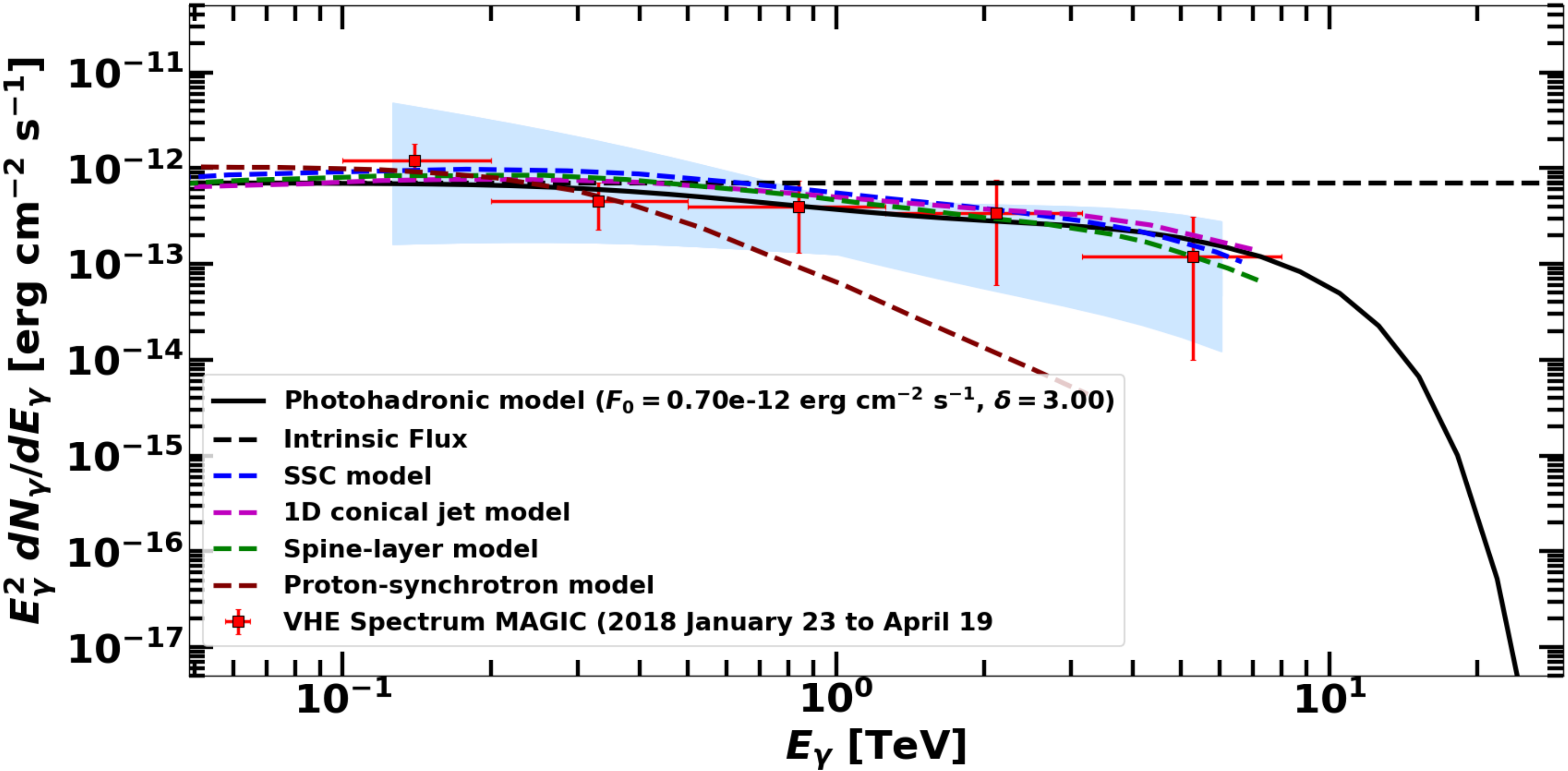}
\caption{\label{fig:figure5}The VHE spectrum of PGC 2402248 is fitted with the leptonic models (One-zone SSC, 1D conical jet, and spine-layer), the proton-synchrotron model and the photohadronic model. The shaded region corresponds to the $1\sigma\ (68.27\%)$ confidence level region for the photohadronic model, where EBL model of~\citet{Franceschini:2008tp} is used for the EBL correction.}
\end{figure*}
\begin{figure*}%figure 6
\includegraphics[width=5.5in]{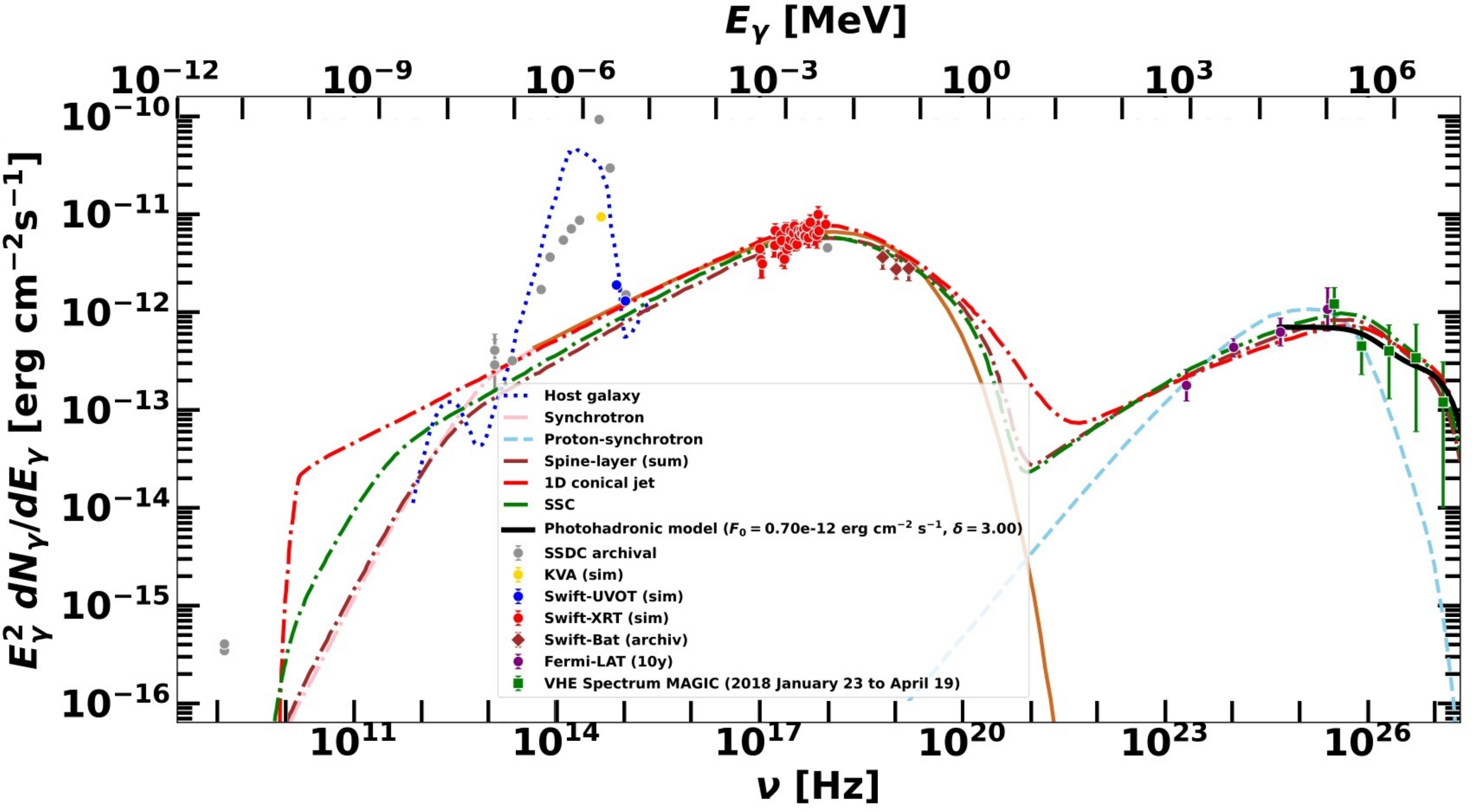}
\caption{\label{fig:figure6}The multiwavelength SED of PGC 2402248 is constructed using the simultaneous multiwavelength observations and the archival data collected by the {\it Fermi}-LAT. The SED is fitted using the leptonic models and the proton-synchrotron model. The photohadronic fit is also shown for comparison.}
\end{figure*}

%%%%%%%%%%%%%%%%%%%%%%%%%%%%%%%%%%%%%%%%
In Ref.~\citet{10.1093/mnras/stz2725}, the multiwavelength SED of PGC 2402248 is fitted using the leptonic and the proton-synchrotron models. It is observed that the one-zone SSC model, the 1D conical jet model, and the spine-layer model (the leptonic models) fit the observed VHE spectrum well as shown in Fig.~\ref{fig:figure5}. Their behavior is similar in the low and the high energy limits. However, the proton synchrotron model does not fit well to the spectrum and is very different from the other fits. Also, its flux falls faster and earlier than in the rest of the models. Although, different leptonic models fit well to the spectrum, among them, the spine-layer model provides the best fit with an estimated $\chi^2_{min}=2.16$ in the VHE region. We compare the photohadronic fit with these models, which is shown in Fig.~\ref{fig:figure5}. In the photohadronic model, the best fit to the spectrum gives $\chi^2_{min}=1.52$. Thus, the  $\chi^2_{min}$ comparison of the photohadronic model and the spine-layer model shows that the photohadronic fit is as good as or better than the leptonic model fits. Also, we observe a slight dip in the spectrum around 1 to 2 TeV energy in the photohadronic models, which is not so obvious in the leptonic scenarios. This happens due to a slight dip in the EBL contribution around this region. We have also shown the multiwavelength SED with the simultaneous data and the archival data from different observations, along with the VHE fit by the leptonic models, the proton-synchrotron model and the photohadronic model in Fig.~\ref{fig:figure6}.
%%%%%%%%%%%%%%%%%%%%%%%%%%%%%%%%%%%%
%%%%%%%%%%%%%%%%%%%%%%%%%%%%%%%%%%%%
The important parameters used in all these models are summarized in Table~\ref{tab:tab1}. By using Eq.~(\ref{eq:epeg}), the $\gamma$-rays in the energy range of $0.1 \, \mathrm{TeV} \le E_{\gamma} \le 8\, \mathrm{TeV}$ correspond to Fermi accelerated proton energy in the range $1 \,\mathrm{TeV} \le E_p \le 80 \, \mathrm{TeV}$. In the inner jet region of radius $R'_f\sim 5\times 10^{15}\, \mathrm{cm}$, to accelerate the protons to energies $E_p\sim 80$ TeV, the magnetic field can be estimated from the relation $eB'=E_p/ R'_f$, which gives $B'\sim 0.5\times 10^{-4}\, \mathrm{G}$. It can be seen from Fig.~\ref{fig:figure6} that the low energy tail of the SSC band starts around  $\epsilon_{\gamma}\simeq 10^{21}\, \mathrm{Hz}$ ($\sim 4.1$ MeV). By using Eq.~(\ref{eq:kingamma}), we can estimate the bulk Lorentz factor $\Gamma$ by assuming that the protons with energy $E_p\simeq 80$ TeV interact with the SSC seed photons with energy $\epsilon_{\gamma}\simeq 4.1$ MeV to produce $E_{\gamma}\simeq 8$ TeV. This gives the maximum value of $\Gamma \simeq 34$. To fit the broad band SED of PGC 2042248, the leptonic models take $\Gamma=30$~\citep{10.1093/mnras/stz2725}. We conclude that our estimate is consistent with the leptonic model value. Using the velocity dispersion measurement, the derived central blackhole mass is $M_{BH}\simeq (4.8 \pm 0.9)\times 10^8\, M_{\odot}$~\citep{10.1093/mnras/staa1144} and this corresponds to Eddington luminosity $L_{Edd}\sim (4.9 - 7.2)\times 10^{46}\, \mathrm{erg\, s^{-1}}$. The integrated VHE flux in the energy range of $0.13\, \mathrm{TeV} \lesssim E_{\gamma} \lesssim 5.3\, \mathrm{TeV}$ gives $F_{\gamma}\sim 4.3\times 10^{-12}\,\mathrm{erg \,cm^{-2}\, s^{-1}}$ and the VHE $\gamma$-ray luminosity is $L_{\gamma}\sim 4.8\times 10^{43}\,\mathrm{erg\, s^{-1}}$.

In order to estimate $\tau_{p\gamma}$ and the photon density in the inner jet region we take the inner jet radius $R'_f\sim 5\times 10^{15}\, \mathrm{cm}$ and the outer jet radius $R'_b\sim 10^{16}\, \mathrm{cm}$~\citep{10.1093/mnras/stz2725}. For a moderate efficiency of the $\Delta$-resonance production we expect $\tau_{p\gamma} < 1$ and this gives $n'_{\gamma,f} < 4\times 10^{11}\, \mathrm{cm^{-3}}$. We can also constrain the value of $\tau_{p\gamma}$ from the fact that the Fermi accelerated proton luminosity (which produces VHE $\gamma$-rays) $L_p=7.5\tau_{p\gamma}^{-1} L_{\gamma}$ should always satisfy $L_p < L_{Edd}/2$. By taking $L_{Edd}\sim  6.0\times 10^{46}\, \mathrm{erg\, s^{-1}}$, we obtain $\tau_{p\gamma} > 0.012$. By taking $\tau_{p\gamma}\sim 0.05$, the SSC photon density in the inner jet region is $n'_{\gamma,f}\sim 2.0\times 10^{10}\, \mathrm{cm^{-3}}$ and the proton luminosity $L_p\sim 7.2\times 10^{45}\, \mathrm{erg\, s^{-1}}$.

In the photohadronic scenario, as previously discussed, the neutrinos produced from the charged pion decay have energy $E_{\nu}=0.5\, E_{\gamma}$. From the VHE flare of PGC 2402248 on April 19, 2018, the maximum observed $\gamma$-ray energy was $E_{\gamma}=8$ TeV, which corresponds to $E_{\nu}=4$ TeV. For a neutrino detector like IceCube, this neutrino energy is very low to be observed. Additionally, in the VHE $\gamma$-ray regime the source is in the low emission state, having the integrated flux $F_{\gamma}\sim 4.3\times 10^{-12}\,\mathrm{erg \,cm^{-2}\, s^{-1}}$. Using Eq.~(\ref{eq:nuflux}), we determine the neutrino flux to be $F_{\nu}\sim 1.6\times 10^{-12}\,\mathrm{erg \,cm^{-2}\, s^{-1}}$, which is also low. Thus, such low energy neutrinos and low neutrino flux pose a challenging detecting task for IceCube.

\section{Discussion and Conclusions}

The MAGIC telescopes for the first time observed multi-TeV $\gamma$-rays from the blazar PGC 2402248 on April 19, 2018. Also, simultaneously, the source was observed in a wide range of frequency bands. The synchrotron peak frequency observed during January - April, 2018 period and the 10 years archival data of {\it Swift}-XRT establishes that PGC 2402248 is a constantly exhibiting extreme HBL. The observed VHE spectrum of the source is flat as compared to several other EHBLs. The broadband SED of PGC 2402248 is fitted using several leptonic models and the proton synchrotron model. It is observed that the leptonic models fit well to the observed VHE spectrum. In previous studies, we have established the success of the photohadronic model in explaining the VHE spectra of several HBLs and EHBLs. For this reason, the VHE spectrum of the EHBL PGC 2402248 is also studied in the context of photohadronic model using different EBL models. It is observed that, the flat VHE spectrum can be fitted very well for the spectral index $\delta=3.0$ which corresponds to a low emission state. The estimated bulk Lorentz factor in this model is consistent with the other leptonic models. Also, we have compared the photohadronic fit with the other leptonic and hadronic fits. We conclude that the photohadronic fit is as good as or better than the other models.

It is to be noted that although the photohadronic model has explained the enigmatic VHE spectra of many HBLs and EHBLs very well, the population of EHBLs detected so far is small and the hard VHE spectra pose a challenge to the leptonic model interpretation. The interpretation of the multi-TeV flaring events in the context of the hadronic models  offers the opportunity to look for high-energy neutrinos by the IceCube neutrino observatory. With the present understanding of the EHBLs and their classification, possibility of other subclass of the EHBL can not be ruled out. For these reasons, it is necessary to undertake future observational programs to look for more EHBLs with simultaneous and/or quasi-simultaneous observations at several wavelengths to elucidate their nature and to test the validity of different emission models. 

%%%%%%%%%%%%%%%%%%%%%%%%%%%%%%%%%%%%%%%%%%%%%%%%%%

\section*{Acknowledgements}
%\addcontentsline{toc}{section}{Acknowledgements}
B. M-C and G. S-C would like to thank CONACyT (México) for partial support. The work of S.S. is partially supported by DGAPA-UNAM (México) Project No. IN103522. Partial support from CSU-Long Beach is gratefully acknowledged. 

%%%%%%%%%%%%%%%%%%%%%%%%%%%%%%%%%%%%%%%%%%%%%%%%%%

\section*{Data Availability}
No new data were generated or analysed in support of this research.

%%%%%%%%%%%%%%%%%%%%%%%%%%%%%%%%%%%%%%%%%%%%%%%%%%

\bibliographystyle{mnras}
\bibliography{PGC2402248}

%%%%%%%%%%%%%%%%%%%%%%%%%%%%%%%%%%%%%%%%%%%%%%%%%%

% Don't change these lines
\bsp	% typesetting comment
\label{lastpage}
\end{document}